\documentstyle[preprint,eqsecnum,tighten,aps]{revtex}

\begin{document}
\draft

\title{Green's matrix from Jacobi-matrix Hamiltonian \thanks{to be published in
Journal of Mathematical Physics}}
\author{B. K\'onya, G. L\'evai, Z. Papp}
\address{Institute of Nuclear Research of the Hungarian \\
Academy of Sciences, \\
P.O. Box 51, H--4001 Debrecen, Hungary}
\date{\today}
\maketitle
\begin{abstract}
We propose two ways for determining the Green's matrix for
problems admitting Hamiltonians that have infinite symmetric
tridiagonal (i.e. Jacobi) matrix form on some basis
representation. In addition to the recurrence
relation comming from the Jacobi-matrix, 
the first approach also requires the matrix elements
of the Green's operator between the first elements of the basis.
In the second approach the recurrence relation is solved
directly by continued fractions and the solution is continued
analytically to the whole complex plane. Both approaches are
illustrated with the non-trivial but calculable example of the
D-dimensional Coulomb Green's matrix. We give the corresponding 
formulas for the D-dimensional harmonic oscillator as well.
\end{abstract}
\vspace{0.5cm}

\pacs{PACS number(s): 02.30.Rz, 02.30.Lt, 03.65.Ge, 02.60.Nm,
 21.45.+v }

\narrowtext
\section{Introduction}

Green's operators play a central role in theoretical physics, especially in
quantum mechanics, since the fundamental equations are formulated as
integral equations containing Green's operators in their kernels. Integral
equation formalisms have an advantage over those based on differential
equations because they automatically incorporate the boundary conditions. In
spite of this fact differential equations are more extensively used in
practical calculations. The reason certainly is that the Green's operators
occurring in integral equations are much more complicated than the
corresponding terms in the Hamiltonian.

A possible way of compensating this drawback is using a representation in
which the Green's operator appears in a simple form. In this respect the
momentum-space representation is rather appealing, as the free Green's
operator is very simple there. This is the main reason why momentum-space
techniques are so frequently used and also why they are capable of coping
with complicated integral equations like the Faddeev equations (see Ref.~%
\cite{gloeckle-report} for a review).

The free Green's operator can also be given analytically between harmonic
oscillator (HO) states \cite{ho1}. This allowed the construction of a
flexible method for solving the Lippmann--Schwinger equation in HO-space,
which contains the free Green's operator in its kernel \cite{ho2}. The
representation of operators on a finite subset of a countable basis, such as
the HO basis, turns the Lippmann--Schwinger equation into a matrix equation.
The completeness of the basis ensures the convergence of the method.
Actually, this is equivalent to a separable expansion of the potential (see
e.g. Ref.~\cite{plessas} for a review). In this approach only the potential
term is approximated, the terms in the Green's operator (the kinetic energy
term in the case of the free Green's operator) are not. Thus, although one
is working with finite matrices, the solution possesses correct asymptotic
behavior.

To give account of the fact that Coulombic asymptotics are genuinely
different from the free one, the kernel of the integral equations describing
Coulombic systems should include Coulombic, rather than free Green's
operators. For the two-body Coulomb Green's operator there exists a
Hilbert-space basis in which its representation is very simple, namely the
Coulomb--Sturmian (CS) basis. In CS-space the Coulomb Green's operator can
be given by simple and well-computable special functions, which can be
continued analytically to the complex plane \cite{papp1}. This is also a
countable basis, so we have a matrix representation.

In the past few years a quantum mechanical approximation method for treating
Coulomb-like interactions in two-body calculations was developed along this
line. The analytic Coulomb Green's matrix allows the extension of the method
to resonant- and scattering-state calculations \cite{papp2}. Since only the
asymptotically irrelevant short-range interaction is approximated, the
correct (two-body) Coulomb asymptotics is guaranteed. The corresponding
computer codes for solving two-body bound-, resonant- and scattering-state
problems were also published \cite{cpc}.

Recently the CS-space representation approach to the Faddeev integral
equations has been applied to solving the three-body bound- and
scattering-state problem with Coulomb interactions \cite{pzwp,pz3sc}. In
this formulation of the equations the most crucial point is calculating the
resolvent of the sum of two independent, thus commuting, two-body Coulombic
Hamiltonians. This is given by the convolution integral \cite{bianchi} 
\begin{equation}
(z-h_1-h_2)^{-1}=\frac {1}{2\pi \mbox{i} } \oint_C\;\mbox{d}%
w\;(z-w-h_1)^{-1} \ (w-h_2)^{-1}.  \label{contourint}
\end{equation}
Here the contour $C$ should encircle, in counterclockwise direction, the
spectrum of $h_2$ without penetrating into the spectrum of $h_1$ . The
analytic nature of the two-body Green's matrix made the evaluation of the
contour integral, also in practice, possible. In fact, the convolution
integral follows directly from the Dunford--Taylor integral representation
of operators \cite{dunford}. A function of an operator $h$ is defined as 
\begin{equation}
f(h)=\frac 1{2\pi \mbox{i}}\oint_C\;\mbox{d}w\;f(w)(w-h)^{-1},
\end{equation}
where $f$ should be analytic on $C$. This way we can calculate complicated
functions of operators as well.

Making use of (\ref{contourint}), we can solve problems which otherwise
would amount to solving non-separable partial differential equations with
unknown boundary conditions. So we believe that the analytic representation
of simple Green's operators is of extreme importance, probably in other
fields of physics too, but certainly for the solution of the underlying
integral equations of quantum mechanics.

Hamiltonians having Jacobi-matrix forms were also extensively studied in the
context of the $L^2$ approach to quantum scattering theory \cite
{l2_approach,l2_g}. Here the Hamiltonian is represented on an appropriate $%
L^2$ basis, which is chosen in such a way, that the asymptotic part of the
Hamiltonian possesses a Jacobi-matrix form. The resulting three-term
recurrence relation can be solved analytically, yielding the expansion
coefficients of both a ''sine-like'' $\widetilde{S}(r)$ and the
''cosine-like'' solution $\widetilde{C}(r)$. The Jacobi-matrix solutions $%
\widetilde{S}(r)$ and $\widetilde{C}(r)$ are then used to obtain the exact
solution to a model scattering problem defined by approximating the
potential $V$ by its projection $V^N$ onto the finite subspace spanned by
the first $N$ basis functions. So, on the level of physical assumptions the $%
L^2$ approach and the methods used in Refs.\ \cite{ho2} and \cite
{papp1,papp2,cpc} are equivalent in the same sense as the Schr\"{o}dinger
equation is equivalent to the Lippmann--Schwinger equation. However, we
believe that the approaches starting from integral equations, especially
those in Refs.\ \cite{papp1,papp2,cpc}, are superior since they allow us to
cope with problems still lacking satisfactory solution, such as three-body
scattering problems with Coulomb interactions. We note that in the $L^2$
approach the Green's function of a Jacobi-matrix Hamiltonian can also be
constructed from the coefficients of the solutions $\widetilde{S}(r)$ and $%
\widetilde{C}(r)$ \cite{l2_g}. However, this construction is applicable only
in very exceptional cases.

In this paper we wish to demonstrate that, if the Hamiltonian appears in a
symmetric infinite tridiagonal, i.e. Jacobi-matrix form in some basis
representation with analytically known matrix elements, then the
corresponding Green's matrix can be given in terms of three-term recurrence
relation. We present two independent methods for determining the Green's
matrix from a three-term recurrence relation.

In our first method (method A) we consider this relation only as a useful
computational tool. In addition to the recurrence relation this approach
also requires,the matrix element of the Green's operator between the first
elements of the basis. This often necessitates the analytical evaluation of
complicated integrals, which restricts its use to exceptional cases only.
Once this matrix element has been calculated, we can resort to the
recurrence relation in order to determine the Green's matrix. However, from
the numerical point of view the recurrence relations can lead to
calculational problems and instabilities \cite{jones}.

In our second approach (method B) we propose direct solution of this
recurrence relation by continued fractions. The richness of the theory of
recurrence relations and continued fractions enable us to avoid the
difficult and strenuous procedure of method A. The inverse of the Green's
matrix can be evaluated solely from the ratio of two successive elements of
the recurrence relation. In method B this ratio is provided by a continued
fraction. This means that, all the above mentioned numerical problems can be
avoided, since the recurrence relation is completely ignored as a
computational tool.

In Sec.~II below we sketch methods A and B. In Sec.~III we illustrate method
A with a non-trivial but calculable example, the D-dimensional Coulomb
problem. The corresponding 3-dimensional formulas had been presented earlier 
\cite{papp1,papp2,cpc} and were extensively used in both two- and three-body
calculations. A summary of the relevant mathematical formulas for continued
fractions and three-term recurrence relations is given in Sec.~IV. In Sec.~V
method B is presented using the example of the D-dimensional Coulomb
problem. This is followed by numerical illustrations in Sec. VI. Finally, in
the Appendix the D-dimensional harmonic oscillator is considered. We show,
that the harmonic oscillator Hamiltonian takes a Jacobi-matrix form on a
harmonic oscillator basis that has a different frequency parameter.

\section{Jacobi-matrix representation}

We first define our representation space for the quantum mechanical problem.
Let us consider the basis states $\{|i\rangle \}$ and $\{|\tilde{i}\rangle \}
$, with $i=0,1,2,\ldots $, which form a complete bi-orthogonal set, i.e. 
\begin{equation}
\langle \tilde{i}|j\rangle =\langle i|\tilde{j}\rangle =\delta _{ij}
\end{equation}
\begin{equation}
{\bf 1}=\sum_{i=0}^\infty |\tilde{i}\rangle \langle i|= \sum_{i=0}^\infty
|i\rangle \langle \tilde{i}|,
\end{equation}
Let us start with the defining equation of the Green's operator $G$
corresponding to Hamiltonian $H$ 
\begin{equation}
{\bf 1}=(E-H)G,  \label{gdef}
\end{equation}
and apply it to the ket $|\tilde{j}\rangle $ 
\begin{equation}
|\tilde{j}\rangle =(E-H)G|\tilde{j}\rangle .
\end{equation}
Inserting a completeness relation between $E-H$ and $G$, and multiplying
form the left by the bra $\langle j|$ we get 
\begin{equation}
\delta _{ij}=\sum_{i^{\prime }=0}^\infty \langle i|(E-H)|i^{\prime }\rangle
\langle \tilde{i^{\prime }}|G|\tilde{j}\rangle .  \label{gginv}
\end{equation}
If $\langle i|(E-H)|i^{\prime }\rangle $ takes a Jacobi-matrix form the
infinite sum is reduced only to three terms and we arrive at a recurrence
relation for the matrix elements $G_{ij}=\langle \tilde{i}|G|\tilde{j}%
\rangle $: 
\begin{equation}
\delta _{ij}=J_{i i-1}G_{i-1 j}+J_{i i}G_{i j}+ J_{i i+1}G_{i+1 j},\ \ \ \ \
i=1,2,\ldots ,j=0,1,\ldots ,  \label{3termg}
\end{equation}
where $J_{ij}=\langle i|(E-H)|j\rangle $ are the elements of the Jacobi-matrix. For the $i=j=0$ case Eq.\ (\ref{gginv}) takes the form 
\begin{equation}
1=J_{01}G_{10}+J_{00}G_{00}.  \label{0rec}
\end{equation}

In method A, if $G_{00}$ is known, we can calculate $G_{10}$ from (\ref{0rec}%
), and then can continue to $G_{j0}$ using Eq.\ (\ref{3termg}).
Interchanging the indices in Eq.\ (\ref{3termg}) we again get a three-term
recurrence relation which can be utilized to generate the $G_{ij}$ elements
from the known $G_{0j}$ terms. The analytic calculation of $G_{00}$ together
with the application of the recurrence relation (\ref{3termg}) with (\ref
{0rec}) constitutes the basic idea of method A.

In method B we compute the Green's matrix without the explicit use of the
recurrence relation. From the theory of special matrices we know that the
inverse of a Jacobi-matrix, i.e. the Green's matrix, possesses the property 
\cite{rozsa} 
\begin{equation}
G_{ij}=\left\{ 
\begin{array}{r@{\quad\mbox{if}\quad}l}
p_iq_j, & i\le j \\ 
p_jq_i, & j\le i
\end{array}
\right. .  \label{rozsapal}
\end{equation}
Therefore, for $j\le N$ and $i\le N$ we can write the resolvent equation (%
\ref{gginv}) as 
\begin{eqnarray}
\delta _{ij} &=&\sum_{i^{\prime }=0}^N (J_{ii^{\prime }}G_{i^{\prime
}i}-\delta _{iN}\ J_{i N+1}\ G_{j N+1})  \nonumber \\
&=&\sum_{i^{\prime }=0}^N(J_{ii^{\prime }}-\delta _{iN}\ \delta _{i^{\prime}
N}\ J_{i N+1}\ p_{N+1}/p_N)G_{i^{\prime } j}  \nonumber \\
&=&\sum_{i^{\prime }=0}^N(J_{ii^{\prime }}-\delta _{iN}\ \delta _{i^{\prime
} N}\ J_{i N+1}\ G_{0 N+1}/G_{0 N})G_{i^{\prime } j} ,  \label{invi}
\end{eqnarray}
i.e. the inverse of the truncated $N\times N$ $G_{ij}^{(N)}$ Green's matrix
is given as 
\begin{equation}
(G_{ij}^{(N)})^{-1}=J_{ij}-\delta _{j N}\ \delta _{i N}\ J_{N N+1}\ G_{0
N+1}/G_{0 N}.  \label{invn}
\end{equation}
Eq. (\ref{invn}) asserts that the inverse of the truncated Green's matrix is
determined by the elements of the Jacobi-matrix and the ratio $G_{0
N+1}/G_{0N}$. This later will be calculated by continued fractions derived
only from the elements of Jacobi-matrix. This is the basic idea of method B.
We notice, that in many practical applications, like the solution of
Lippmann--Schwinger equations, it is directly the $(G_{j,i}^{(N)})^{-1}$
that is really needed (see e.g.\ Ref.\ \cite{pzwp,pz3sc}).

\section{D-dimensional Coulomb Green's matrix in Coulomb--Sturmian
representation}

\label{coul}

Here we define the Coulomb--Sturmian basis and show that on this particular
basis the D-dimensional Coulomb Hamiltonian has a Jacobi-matrix structure. 
We also derive the corresponding three-term recurrence relation for the
Green's matrix and present its solution via method A. In doing so we apply
techniques established already for the three-dimensional ($D=3$) case \cite
{papp1}.

Let us consider the radial Schr\"odinger equation for the D-dimensional
hydrogen atom in the $l$'th partial wave 
\begin{equation}
H^C \psi(r)\equiv \left[ -\frac{\hbar^{2}}{2m} \left( \frac{\mbox{d}^2 }{%
\mbox{d} r^2} - \frac{1}{ r^2}\left(l+\frac{D-3}{ 2}\right) \left(l+\frac{%
D-1}{ 2}\right) \right) - \frac{ Z}{r} \right] \psi(r)=E\psi(r).  \label{hh0}
\end{equation}
(See, e.g. Ref. \cite{knt85} and references.) The bound-state energy
spectrum is given by 
\begin{equation}
E_{n_r l}= -\frac{mZ^2}{ 2\hbar^2(n_r+l+\frac{D-1}{ 2})^2},  \label{hen}
\end{equation}
and the corresponding wave functions are 
\begin{equation}
\psi_{n_r l}(r)= a_0 \left(\frac{r_0 \Gamma(n_r+1)} {
2\Gamma(n_r+2l+D-1)}\right)^{1/2} \exp(-\frac{a_0}{2}r) (a_0 r)^{l+\frac{D-1%
}{2}} L_{n_r}^{(2l+D-2)}(a_0 r),  \label{hwf}
\end{equation}
where we used the notation $a_0= ((n_r+l+\frac{D-1}{2})r_0)^{-1}$ and $r_0= {%
\hbar^2}/{2mZ}$.

The Coulomb--Sturmian equation has a structure similar to the eigenvalue
equation (\ref{hh0}) 
\begin{equation}
\left(-\frac{\mbox{d}^2 }{\mbox{d} r^2} + \frac{1}{ r^2}\left(l+\frac{D-3%
}{ 2}\right)\left(l+\frac{D-1}{ 2}\right) -\frac{(2n+2l+D-1)b_{S}}{ r}
+b_{S}^2\right) \phi(b_{S},r)=0  \label{cse}
\end{equation}
and is solved by the Coulomb--Sturmian (CS) functions 
\begin{equation}
\langle r\vert nl\rangle \equiv \phi_{n l}(b_{S},r) = \left( \frac{%
\Gamma(n+1)}{ \Gamma(n+2l+D-1)}\right)^{1/2} \exp(-b_{S}r) (2b_{S}r)^{l+%
\frac{D-1}{2}} L_n^{(2l+D-2)}(2b_{S}r),  \label{csf}
\end{equation}
which are the generalizations of the corresponding CS functions for the
three-dimensional case \cite{rotenberg}. Here $b_{S}$ is a real parameter,
thus we have the bound-state CS functions.

Introducing the notation $\langle r\vert \widetilde{ nl} \rangle \equiv
\phi_{n l}(b_{S},r)/r$ we can express the orthogonality and completeness of
the CS functions as 
\begin{equation}
\langle nl \vert \widetilde{ n^{\prime}l} \rangle =\delta_{nn^{\prime}}
\label{csort}
\end{equation}
and 
\begin{equation}
{\bf 1}=\sum _{n=0}^\infty | \widetilde{nl} \rangle \langle nl |
=\sum_{n=0}^\infty | nl \rangle \langle \widetilde{nl} |,
\end{equation}
respectively, confirming that they form a bi-orthonormal basis.

The overlap of two CS functions can be written in terms of a three-term
expression 
\begin{eqnarray}
\langle nl|n^{\prime }l\rangle &=&(2b_S)^{-1}\left[ \delta_{nn^{\prime
}}(2n+2l+D-1)-\delta _{nn^{\prime}-1} ((n+1)(n+2l+D-1))^{1/2}\right. 
\nonumber \\
&&\left. -\delta _{nn^{\prime }+1}(n(n+2l+D-2))^{1/2}\right] .  \label{cs3t}
\end{eqnarray}
A similar expression holds for the matrix element of $H^C$: 
\begin{eqnarray}
\langle nl|H^C|n^{\prime }l\rangle &=&\frac{\hbar ^2b_S}{4m}\left[
\delta_{nn^{\prime }}\left( 2n+2l+D-1- \frac{2}{r_0b_S}\right) \right. 
\nonumber \\
&&\left. +\delta _{nn^{\prime }-1}((n+1)(n+2l+D-1))^{1/2}+\delta
_{nn^{\prime }+1}(n(n+2l+D-2))^{1/2}\right].  \label{csh0cs2}
\end{eqnarray}
>From the above two equations follows the Jacobi-matrix structure of $J^C_{n
n^{\prime}}=\langle nl| E-H^C |n^{\prime }l\rangle$.

The Jacobi-matrix structure immediately implies a three-term recurrence
relation (\ref{3termg}) which, in the method A, can only be solved if $%
G^C_{0 0}$ is at our disposal. This matrix element can be gained by
evaluating an integral of the D-dimensional Coulomb--Green's function 
\begin{equation}
G_{lD}^{C}(r,r^{\prime},E)= -\frac{\mbox{i}m}{\hbar^2 k} \frac{\Gamma(l+%
\frac{D-1}{2}+ \mbox{i}\gamma)} {\Gamma(2l+D-1)}{\cal M}_{-\mbox{i}%
\gamma, l+\frac{D}{2}-1} (-2\mbox{i}kr_<){\cal W}_{-\mbox{i}\gamma, l+%
\frac{D}{2}-1}(-2\mbox{i} kr_>),  \label{grrvk}
\end{equation}
with the $n=n^{\prime}=0$ CS functions (\ref{csf}), where $\gamma=Z/(2r_0k)$
and $k$ is the wave number. Using the formula by Buchholz \cite{buchholz} 
\begin{eqnarray}
&&\Gamma\left(\frac{\mu+1}{2}-\kappa\right){\cal W}_{\kappa, \frac{\mu}{2}%
}(a_1t) {\cal M}_{\kappa,\frac{\mu}{2}}(a_2t)  \nonumber \\
&&=t(a_1 a_2)^{\frac{1}{2}}\int_0^{\infty} \exp\left(-\frac{1}{2}(a_1+a_2)t
\cosh w\right)I_{\mu}(t(a_1 a_2)^{\frac{1}{2}}\sinh w) \left(\coth \frac{w}{2%
}\right)^{2\kappa} \mbox{d}w  \label{wm}
\end{eqnarray}
the integration can be performed analytically \cite{zon,papp1} and the final
result is 
\begin{eqnarray}
G^C_{0 0} &=&-\frac{4mb_S}{\hbar^2(b_S-\mbox{i}k)^2} \frac{1}{l+(D-1)/2+%
\mbox{i}\gamma}  \nonumber \\
&\times&\;_2F_1\left(-l-\frac{D-3}{2}+\mbox{i}\gamma, 1;l+\frac{D+1}{2}+%
\mbox{i}\gamma+2; \left(\frac{b_S+\mbox{i}k} {b_S-\mbox{i}%
k}\right)^2\right).  \label{g00}
\end{eqnarray}

\section{Continued fractions and three-term recurrence relations}

Based on the mathematical literature \cite{lorentzen,jones} we give a brief
review of the underlying mathematical theorems. Let $\{a_n(z)\}_1^\infty$
and $\{b_n(z)\}_0^\infty$, $a_n(z)\neq 0$, be two sequences of complex
valued functions defined on the region $D$ of the complex plane. We define
the linear fractional transformation $s_n(w,z)$ as 
\begin{equation}
s_n(w_n,z)=\frac{a_n(z)}{b_n(z)+w_n},\quad n\geq 1,\qquad
s_0(w_0,z)=b_0(z)+w_0,
\end{equation}
and 
\begin{equation}
S_n(w_n,z)=S_{n-1}(s_n(w_n,z),z),\qquad S_0(w_0,z)=s_0(w_0,z).
\end{equation}
A continued fraction is an ordered pair 
\begin{equation}
((\{a_n(z)\},\{b_n(z)\}),\{f_n(z)\}),
\end{equation}
where $\{f_n(z)\}$ is given by 
\begin{equation}
f_n(z)=S_n(0,z), \qquad n=0,1,2,3,\ldots.
\end{equation}
Here $S_n(w_n,z)$ is called the $n$th approximant of the continued fraction
with respect to the $\left\{ w_n\right\} _{n=0}^\infty $ complex series. $%
S_n(w_n,z)$ can be written, using one of the standard notations, as 
\begin{equation}
S_n(w_n,z)=b_0(z)+\frac{a_1(z)}{b_1(z)}%
{ \atopwithdelims.. +}
\frac{a_2(z)}{b_2(z)}
{ \atopwithdelims.. +\cdots +}
\frac{a_n(z)}{b_n(z)+w_n}.
\label{continued1}
\end{equation}
The convergence of a continued fraction means the convergence of the
sequence of approximants $S_n(w_n,z)$ to an extended complex number 
\begin{equation}
f(z)=\lim_{n\rightarrow \infty }S_n(w_n,z)= b_0(z)+K_{n=1}^\infty \left( 
\frac{a_n(z)}{b_n(z)}\right),  \label{cfdef2}
\end{equation}
where 
\begin{equation}
K_{n=1}^\infty \left( \frac{a_n(z)}{b_n(z)}\right)= 
\frac{a_1(z)}{b_1(z)}%
{ \atopwithdelims.. +}
\frac{a_2(z)}{b_2(z)}%
{ \atopwithdelims.. +\cdots +}
\frac{a_n(z)}{b_n(z)} {%
{  \atopwithdelims.. +\cdots}
}.
\end{equation}
It should be noted that if $f(z)$ exists for two different sequences of $%
\{\omega_n\}$ then $f(z)$ is unique.

A special class of continued fractions for which the limits 
\begin{equation}
\lim_{n\rightarrow \infty }a_n(z)=a(z)\qquad \text{and}\qquad
\lim_{n\rightarrow \infty }b_n(z)=b(z)
\end{equation}
exist for all $z\in D$ is called limit 1-periodic continued fractions. The
fixed points $w_{\pm }(z)$ of the linear fractional transformation 
\begin{equation}
s(w,z)=\lim_{n\rightarrow \infty }s_n(w_n,z)=\frac{a(z)}{b(z)+w}, \quad
\end{equation}
where $w=\lim_{n\rightarrow \infty }w_n$, are given as the solution of the
quadratic equation 
\begin{equation}
w=\frac{a(z)}{b(z)+w},  \label{fixed}
\end{equation}
\begin{equation}
w_{\pm }(z)=-b(z)/2 \pm \sqrt{(b(z)/2)^2+a(z)}.
\end{equation}
The $w_{\pm }(z)$ with smaller modulus is called attractive fixed point,
while the other one is called as repulsive fixed point. Since $w_{\pm}(z)$
represent the tail of a limit 1-periodic continued fraction we can speed up
the convergence using the attractive fixed point in the approximant $S_n(w,z)
$.

The idea of the analytic continuation of the continued fraction $f(z)$ in
Eq.\ (\ref{cfdef2}) is based on the proper choice of $\{w_n\}$ in the
approximant $S_n(w_n,z)$ \cite{lorentzen,waadeland}. 
If a continued fraction exists in a certain
complex region $z\in D$ then in many cases it is possible to extend the
region of convergence to a larger domain $D^{*}\supseteq D$, where $D^{*}$
depends  on the choice of the functions $w_n(z)$. In the case of limit
1-periodic continued fractions the analytic continuation is defined with the
help of the fixed points $w_{\pm }(z)$ of Eq.\ (\ref{fixed}) as 
\begin{equation}
f_{D^{*}}(z)=\lim\limits_{n\to \infty }S_n(w_{\pm }(z),z).  \label{analfolyt}
\end{equation}

The computation of the approximants $S_n(w_{\pm }(z),z)$ might be unstable
for certain $z$, which leads to unsatisfactory convergence. This problem can
be overcome by using the Bauer--Muir transformation \cite{lorentzen}. The
Bauer--Muir transform of a continued fraction $b_0(z)+K\left(a_n(z)/b_n(z)%
\right) $ with respect to a sequence of complex numbers $\left\{ w_n\right\}
_{n=0}^\infty $ is the continued fraction $d_0(z)+K\left(
c_n(z)/d_n(z)\right) $, whose ``classical'' approximants $S_n(0,z)$ are
equal to the modified approximants $S_n(w_n,z)$ of the original continued
fraction. The transformed continued fraction exists and can be calculated as 
\begin{eqnarray}
d_0 &=&b_0+w_0,\quad c_1=\lambda _1,\quad d_1=b_1+w_1,  \label{bauermuir} \\
c_i &=&a_{i-1}q_{i-1,}\quad d_i=b_i+w_i-w_{i-2}q_{i-1}, \quad i\geq 2, 
\nonumber \\
\lambda _i &=&a_i-w_{i-1}(b_i+w_i),\quad q_i= \lambda _{i+1}/\lambda _i\quad
i\geq 1,  \nonumber
\end{eqnarray}
if and only if $\lambda _i\neq 0$ for $i=1,2,\ldots .$

We now return to the three-term recurrence relation focusing on their
intimate relations to continued fractions. A three-term recurrence relation
can be written as 
\begin{equation}
X_{n+1}=b_nX_n+a_nX_{n-1}\qquad n=1,2,3,\ldots ,  \label{3term}
\end{equation}
where $a_n,b_n$ are complex numbers and $a_n\neq 0$. The solutions of a
three-term recurrence relation span a two-dimensional linear space. The $%
\left\{ x_n\right\} $ nontrivial (i.e.$\neq \left\{ 0\right\} $) solution is
said to be minimal if there exists another solution $\left\{ y_n\right\} $
such that 
\begin{equation}
\lim_{n\rightarrow \infty }x_n/y_n=0.
\end{equation}
Solution $\left\{ y_n\right\} $ here is called dominant. The minimal
solution is unique, apart from a multiplicative constant.

The existence of the minimal solution is strongly related to the convergence
of a continued fraction constructed from the coefficients of the recurrence
relation. This connection is revealed by Pincherle's theorem \cite
{jones,lorentzen}. 
According to this the following statements hold: \\A: (%
\ref{3term}) has a minimal solution if and only if the continued
fraction 
\begin{equation}
K_{n=1}^\infty \left(\frac{a_n}{b_n}\right)=\frac{a_1}{b_1}%
%TCIMACRO{\QATOPD. . {}{+} }
%BeginExpansion
{ \atopwithdelims.. +}
%EndExpansion
\frac{a_2}{b_2}%
%TCIMACRO{\QATOPD. . {}{+\cdots +} }
%BeginExpansion
{ \atopwithdelims.. +\cdots +}
%EndExpansion
\frac{a_n}{b_n}%
%TCIMACRO{\QATOPD. . {}{+\cdots } }
%BeginExpansion
{ \atopwithdelims.. +\cdots }
%EndExpansion
\label{pincherle1}
\end{equation}
converges, \\B: if $\{X_n\}$ is a minimal solution then for
$N=0,1,2,\ldots, $
\begin{equation}
\frac{x_{N+1}}{x_N}=-K_{n=1}^\infty \left( 
\frac{a_{n+N}}{b_{n+N}}\right) =-%
\frac{a_{1+N}}{b_{1+N}}%
%TCIMACRO{\QATOPD. . {}{+} }
%BeginExpansion
{ \atopwithdelims.. +}
%EndExpansion
\frac{a_{2+N}}{b_{2+N}}%
%TCIMACRO{\QATOPD. . {}{+\cdots +} }
%BeginExpansion
{ \atopwithdelims.. +\cdots +}
%EndExpansion
\frac{a_{n+N}}{b_{n+N}}%
%TCIMACRO{\QATOPD. . {}{+\cdots } }
%BeginExpansion
{ \atopwithdelims.. +\cdots }
%EndExpansion
.  \label{pincherle2}
\end{equation}
The second statement asserts that the ratio of two successive element of the
minimal solution is provided by a continued fraction.

\section{Continued fraction for $G_{0 N+1}/G_{0 N}$}

First we show that in certain domain of the complex plane the physical
relevant solution of the recurrence relation (\ref{3termg}) for the Green's
matrix is the minimal solution. In case of short-range potentials the
Green's function can be constructed as \cite{newton} 
\begin{equation}
G(r,r^\prime,k)=\varphi_l(k,r_<) f_l^{(+)}(k,r_>)/{\cal F}(k),  \label{g}
\end{equation}
where $\varphi_l(k,r)$ is the regular solution, ${\ f}_l^{(+)}(k,r)$ is the
Jost solution, ${\cal F}(k)$ is the Jost function and $k$ is the wave
number. The Jost solution is defined by the relation 
\begin{equation}
\lim\limits_{r\to\infty} \mbox{e}^{\mp i kr} {\ f}_l^{(\pm)}(k,r)=1.
\label{jostdef}
\end{equation}
Let us define a ``new'' Green's function as 
\begin{equation}
\widetilde{G}(r,r^\prime,k)=\varphi_l(k,r_<) f_l(k,r_>)/{\cal F}(k),
\end{equation}
where $f_l$ is a linear combination of ${\ f}_l^{(+)}$ and ${\ f}_l^{(-)}$.
If $\Re E < 0$ ${\ f}_l^{(+)}$ is exponentially decreasing and ${\ f}_l^{(-)}
$ is exponentially increasing. Thus, for any $\widetilde{G}$ we have 
\begin{equation}
\lim\limits_{r^\prime \to\infty} \frac{G(r,r^\prime,k)} {\widetilde{G}%
(r,r^\prime,k)} =0, \qquad \mbox{if\ \ } \Re E < 0.  \label{gto0}
\end{equation}
We note, that both $G$ and $\widetilde{G}$ satisfy the defining equation
Eq.\ (\ref{gdef}), but only $G$ of Eq.\ (\ref{g}) is the physical Green's
function. The above considerations, with a slight modification in Eq.\ (\ref
{jostdef}), are also valid for the Coulomb case.

An interesting result of the study of Ref.\ \cite{l2_g} is that the Green's
matrix from Jacobi-matrix Hamiltonian, in correspondence with (\ref{rozsapal}%
), has an analogous structure to Eq.\ (\ref{g}) 
\begin{equation}
G_{ii^\prime}(k)=({\varphi_l})_{i_<}(k) (f_l^{(+)})_{i_>}(k)/ {\cal F}(k),
\label{gJ}
\end{equation}
where $({\varphi_l})_{i}(k)=\langle \varphi_l(k) \vert \widetilde{i} \rangle$
and $(f_l^{(+)})_{i}(k)=\langle f_l^{(+)}(k) \vert \widetilde{i} \rangle$.
Similarly, we define $(f_l)_{i}(k)=\langle f_l(k) \vert \widetilde{i} \rangle
$ and 
\begin{equation}
\widetilde{G}_{ii^\prime}(k)= ({\varphi_l})_{i_<}(k) (f_l)_{i_>}(k)/ {
\cal F}(k).  \label{gJt}
\end{equation}
On the $\Re E < 0$ region of the complex plane as $r\to\infty$ $f_l (k,r)$
exponentially dominates over $f_l^{(+)}(k,r)$, thus for their $L^2$
representation the following relation holds 
\begin{equation}
\lim\limits_{i \to\infty} \frac{ (f_l^{(+)})_{i}(k)} {(f_l)_{i}(k)} =0,
\qquad \mbox{if\ \ } \Re E < 0.  \label{fgto0}
\end{equation}
This implies a similar relation for the Green's matrices 
\begin{equation}
\lim\limits_{i^\prime \to\infty} \frac{G_{ii^\prime}(k)} {\widetilde{G}%
_{ii^\prime}(k)} =0, \qquad \mbox{if\ \ } \Re E < 0.  \label{ggto0}
\end{equation}

So, in the $\Re E<0$ region of complex $E$-plane the physical relevant
Green's matrix $G_{ii^{\prime }}$ appears as the minimal solution of
recurrence relation (\ref{3termg}). Thus, according to Pincherle's theorem (%
\ref{pincherle2}), the ratio needed in Eq.\ (\ref{invn}) for the Green's
matrix can be calculated by the continued fraction 
\begin{equation}
\frac{G_{0N+1}(\epsilon )}{G_{0N}(\epsilon )}=-K_{i=N}^\infty \left( \frac{%
a_i}{b_i}\right) ,  \label{boundrec}
\end{equation}
where $a_i=-J_{ii-1}/J_{ii+1}$, $b_i=-J_{ii}/J_{ii+1}$ and $J_{ij}$ is the
Jacobi-matrix.

In the case of D-dimensional Coulomb Green's matrix we have 
\begin{equation}
a_i=-\sqrt{\frac{i(i+2l^{^{\prime }}+1)}{ (i+1)(i+2l^{^{\prime }}+2)}}\qquad
i=1,2,\ldots ,  \label{ehatok}
\end{equation}
\begin{equation}
b_i(\epsilon )= \frac{2\left( \epsilon -b_S^2\right) (i+l^{^{\prime
}}+1)-2b_S Z^{\prime }}{\left( \epsilon +b_S^2\right) \sqrt{
(i+1)(i+2l^{^{\prime }}+2)}}\qquad i=0,1,2,\ldots
\end{equation}
with $\epsilon =2mE/\hbar ^2$, $Z^{\prime }=2mZ/\hbar^2$ and $l^{^{\prime
}}=l+(D-3)/2$.

On the region of scattering states the recurrence relation does not have
minimal solution and the continued fraction (\ref{boundrec}) diverges. On
the other hand, ${G_{0 N+1}(\epsilon )}/{G_{0 N}(\epsilon )}$ is an analytic
function of $\epsilon $, and there is a region of the complex plane where we
have a representation for this function, thus values on other regions can be
obtained by the analytic continuation of the bound-state formula, i.e. by
the analytic continuation of the continued fraction (\ref{boundrec}). Since
now we have a limit 1-periodic continued fraction this, according to Eq.\ (%
\ref{analfolyt}), can be achieved with fixed points 
\begin{equation}
w_{\pm }(\epsilon )=-b(\epsilon )/2\pm \sqrt{(b(\epsilon))^2/4+a}
\label{coulombfix}
\end{equation}
where 
\begin{eqnarray}
a &=&\lim_{i\rightarrow \infty } a_i=-1 \\
b (\epsilon ) &=&\lim_{i\rightarrow \infty } b_i= 2(\epsilon
-b_S^2)/(\epsilon +b_S^2).  \nonumber
\end{eqnarray}

Considering the formula for Green's operators \cite{taylor} 
\begin{equation}
\langle \widetilde{i} \vert G(E+\mbox{i}0)\vert \widetilde{i} \rangle -
\langle \widetilde{i} \vert G(E-\mbox{i}0) \vert \widetilde{i} \rangle =
-2\pi \mbox{i} \langle \widetilde{i} \vert \psi(E) \rangle \langle \psi(E)
\vert \widetilde{i} \rangle,
\end{equation}
where $\psi(E)$ is the scattering wave function, and the analytic properties
of Green's operators we can readily derive that the imaginary part of $%
\langle \widetilde{i} \vert G(E+\mbox{i}0)\vert \widetilde{i} \rangle$
should be negative. This condition can only be fulfilled with the choice of $%
w_{+}$. This choice gives an analytic continuation to the physical sheet,
while $w_{-}$, which also converges, gives an analytic continuation to the
unphysical sheet.

>From the above considerations it follows that utilizing the Jacobi-matrix
only the Green's matrix can be obtained for arbitrary complex energies by
simply evaluating a continued fraction.

\section{Numerical illustrations}

Below we demonstrate the convergence and the numerical accuracy of method B.
We calculate the matrix element $G_{00}^C(\epsilon )$ of D-dimensional
Coulomb Green's operator for the $l=0$ and $D=3$ case at bound- and
scattering-state energies. We examine the convergence of continued fraction
with different choice of $\{w_n\}$ in Eq.\ (\ref{cfdef2}) and the effect of
Bauer-Muir transformation. For comparison we also give the exact value for $%
G_{00}^C(\epsilon )$ (\ref{g00}).

In case of $\Re \epsilon \le 0$ we take the $w_n=0$, $w_n=w_{+}$ and $%
w_n=w_{-}$ choices. In Table I we can observe excellent convergence to the
exact value in all cases. The choice of $w_n$ influences only the speed of
convergence.

In the region of $\Re \epsilon \ge 0$, in complete accordance with
Pincherle's theorem, the continued fraction (\ref{boundrec}) diverges, only
its analytic continuation with $w_n=w_{+}$ and $w_n=w_{-}$ is convergent.
However, as the first column in Table II shows, the convergence is rather
poor. This can considerably be improved by the repeated application of
Bauer-Muir transforms. In fact, an accuracy similar to the bound-state case
can easily be reached here with e.g.\ an eightfold Bauer-Muir transform.

In order that we have a more stringent test we have performed the contour
integral 
\begin{equation}
I(C)=\frac 1{2\pi i}\oint_C\;\mbox{d}\epsilon \;G_{00}(\epsilon ).
\label{contourint2}
\end{equation}
If the domain surrounded by $C$ does not contain any pole, then $I(C)\equiv 0
$. If this domain contains a single bound-state pole, then $I(C)=\langle 
\widetilde{0}|\psi \rangle \langle \psi | \widetilde{0}\rangle $ must hold,
while if $C$ circumvents the whole spectrum then $I(C)=\langle \widetilde{0}|%
\widetilde{0}\rangle $ is expected. With appropriate selection of Gauss
integration points we could reach 12 digits accuracy in all cases. This
demonstrates that the calculation of the Green's matrices from J-matrices
via continued fractions is accurate on the whole complex plane.

\section{Summary and conclusions}

In this paper we have shown that if in some basis representation the
Hamiltonian takes a Jacobi-matrix form the corresponding Green's matrix can
be calculated on the whole complex energy plane by a continued fraction,
whose coefficients are related to the elements of the Jacobi-matrix. To
justify this statement we presented the example of the D-dimensional Coulomb
problem, in particular, we calculated the Coulomb--Sturmian-space
representation of the D-dimensional Coulomb Green's operator. Numerical
examples proved the accuracy and the efficiency of the method.

The applicability of the techniques presented here can be extended beyond
the examples discussed above. We may have a physical situation in which only
the asymptotic part of the Hamiltonian has Jacobi-matrix structure. In this
case, like in the $L^2$ approach we can approximate $V$ by its projection $%
V^N$ onto a finite subspace spanned by the first $N$ basis states and can
generate the analytic Green's matrix as a solution of a Lippmann--Schwinger
matrix equation (see e.g.\ in Ref.~\cite{pzwp,pz3sc}). 
Also, our analytic two-body
Green's matrices may be used to derive Green's matrices of composite systems
via convolution integrals based on two-body problems.

It should be emphasized that we used the Jacobi-matrix form of a particular
Hamiltonian only and the method is applicable to any Jacobi-matrix
Hamiltonian if the matrix elements are know analytically. This later
requirement may be relaxed and thus we can determine approximate Green's
matrices from J-matrices generated by the numerical Lanczos procedure.

\acknowledgments

Authors are indebted to K.~F.~P\'al for his contribution to the early stages
of this work. This work has been supported by the OTKA contracts No.\ T17298
and \ F20689.

\section{Appendix}

Here we present the formulas analogous to those in Section \ref{coul} for
the D-dimensional Harmonic oscillator. The radial Schr\"odinger equation in
this case is 
\begin{equation}
H^{HO}\psi(r)\equiv \left[ -\frac{\hbar^2}{2m}\left( \frac{\mbox{d}^2 }{%
\mbox{d} r^2} -\frac{1}{ r^2}\left(l+\frac{D-3}{ 2}\right)\left( l+\frac{%
D-1}{ 2}\right)\right) +\frac{1}{2}m\omega^2 r^2 \right] \psi(r)=E\psi(r).
\label{hh0o}
\end{equation}
The energy eigenvalues are 
\begin{equation}
E_{n l}=\hbar\omega\left(2n +l+\frac{D}{2}\right)  \label{heno}
\end{equation}
and the corresponding wave functions can be written as 
\begin{equation}
\langle r \vert \omega,nl \rangle \equiv \psi_{n l}(\omega,r) = v^{\frac{1}{4%
}}\left(\frac{2\Gamma(n +1)} { \Gamma(n +l+\frac{D}{2})}\right)^{1/2} \exp(-%
\frac{v}{2}r^2) (vr^2)^{\frac{l}{2}+\frac{D-1}{4}} L_{n} ^{(l+\frac{D}{2}%
-1)}(vr^2),  \label{hwfo}
\end{equation}
where $v=m\omega /\hbar$. The harmonic oscillator functions are orthonormal
and form a complete set in the usual sense.

The harmonic oscillator Hamiltonian with parameter $\omega $ on the basis of
harmonic oscillator function with parameter $\omega ^{\prime }$ takes a
Jacobi-matrix form 
\begin{eqnarray}
\langle \omega ^{\prime },nl|H^{HO}(\omega,E)| \omega ^{\prime },n^{\prime
}l\rangle & = & \delta_{n n^{\prime }} \left( \hbar \frac{\omega ^2+{\omega
^{\prime }}^2}{2\omega ^{\prime }}\left(2n^{\prime }+l+\frac D2\right)\right)
\nonumber \\
&& - \delta_{n n^{\prime }-1 } \hbar \frac{\omega ^2-{\omega ^{\prime }}^2}
{2\omega^{\prime }}\left( n^{\prime }\left(n^{\prime }+l+ \frac
D2-1\right)\right) ^{1/2}  \nonumber \\
&& - \delta_{n n^{\prime }+1 } \hbar \frac{\omega ^2-{\omega ^{\prime }}^2}
{2\omega^{\prime }}\left( (n^{\prime }+1)\left(n^{\prime }+l +\frac
D2\right)\right) ^{1/2}  \label{hog0ho}
\end{eqnarray}

The calculation of the Green's matrix via method B goes analogously to the
Coulomb case. The matrix element $\langle 0l|G^{HO}(E )|0l\rangle$, which
can be used in method A, is given as 
\begin{eqnarray}
\langle 0l|G^{HO}(E )|0l\rangle &=&-8\frac{\omega ^{\prime } \omega }{{%
\omega ^{\prime }}^2+\omega ^2}\frac 1{(E-\hbar \omega (l+\frac D2))} 
\nonumber \\
&\times &\ _2F_1\left( -\frac l2-\frac D4+1-\frac E{2\hbar \omega }, 1;\frac
l2+\frac D4+1-\frac E{2\hbar \omega };\left( \frac{\omega -\omega ^{\prime }%
}{\omega +\omega ^{\prime }}\right) ^2\right).  \label{g00o}
\end{eqnarray}

\begin{table}[tbp]
{\small 
\begin{tabular}{r|ccc|}
& \multicolumn{3}{c|}{$\epsilon=(-100,0)$} \\ 
$n$ & $G_{B}^{(0)}$ & $G_{B}^{(w_{+})}$ & $G_{B}^{(w_{-})}$ \\ \hline
1 & (-5.44922314793965,0) & (-5.59142801316938,0) & (-0.92906408986331,0) \\ 
2 & (-5.54075476366523,0) & (-5.56131039101044,0) & (-4.70080363351349,0) \\ 
3 & (-5.55501552420656,0) & (-5.55812941271530,0) & (-5.39340957492282,0) \\ 
4 & (-5.55726787304507,0) & (-5.55775017508704,0) & (-5.52662100417805,0) \\ 
5 & (-5.55762610832912,0) & (-5.55770176067796,0) & (-5.55192403535264,0) \\ 
6 & (-5.55768333962797,0) & (-5.55769530213874,0) & (-5.55663951922343,0) \\ 
7 & (-5.55769251168083,0) & (-5.55769441374319,0) & (-5.55750389878413,0) \\ 
8 & (-5.55769398510141,0) & (-5.55769428874846,0) & (-5.55766025755765,0) \\ 
9 & (-5.55769422223276,0) & (-5.55769427085427,0) & (-5.55768824220786,0) \\ 
10 & (-5.55769426045319,0) & (-5.55769426825710,0) & (-5.55769320762502,0)
\\ 
11 & (-5.55769426662100,0) & (-5.55769426787592,0) & (-5.55769408236034,0)
\\ 
12 & (-5.55769426761735,0) & (-5.55769426781946,0) & (-5.55769423553196,0)
\\ 
13 & (-5.55769426777843,0) & (-5.55769426781103,0) & (-5.55769426221577,0)
\\ 
14 & (-5.55769426780450,0) & (-5.55769426780976,0) & (-5.55769426684377,0)
\\ 
15 & (-5.55769426780872,0) & (-5.55769426780957,0) & (-5.55769426764335,0)
\\ 
16 & (-5.55769426780940,0) & (-5.55769426780954,0) & (-5.55769426778103,0)
\\ 
17 & (-5.55769426780951,0) & (-5.55769426780954,0) & (-5.55769426780465,0)
\\ 
18 & (-5.55769426780954,0) &  & (-5.55769426780870,0) \\ 
19 & (-5.55769426780954,0) &  & (-5.55769426780939,0) \\ 
20 &  &  & (-5.55769426780950,0) \\ 
21 &  &  & (-5.55769426780954,0) \\ 
22 &  &  & (-5.55769426780954,0) \\ \hline
& \multicolumn{3}{c|}{$G_{A} =(-5.55769426780954, 0)$}
\end{tabular}
}
\caption{Convergence of the continued fraction for first element of the
Green's matrix at $\Re \epsilon<0$ with method B. The first, second and
third column contain approximants of the continued fraction with $w_n=0$, $%
w_n=w_+$ and $w_n=w_-$, respectively. For comparison we also give the exact
result. All the $G$ values are scaled with $10^2$.}
\label{kotott}
\end{table}

\begin{table}[tbp]
\label{szoras} {\small 
\begin{tabular}{r|cccc|}
& \multicolumn{4}{c|}{$\epsilon=(1000,1)$} \\ 
$n$ & $G_{B}^{(w_{+})}(0)$ & $G_{B}^{(w_{+})}(1) $ & $G_{B}^{(w_{+})}(5) $ & 
$G_{B}^{(w_{+})}(8) $ \\ \hline
1 & (1.076,-0.678) & (1.8072,-0.3293) & (-0.4129544,-0.14238595) & 
(-0.2321154,-0.073120618) \\ 
5 & (1.074,-0.279) & (1.1225,-0.3783) & (4.29352799,-1.63424931) & 
(-1.4408861,-0.350899497) \\ 
10 & (1.198,-0.325) & (1.1425,-0.3162) & (1.13445656,-0.32244006) & 
(1.20667672,0.237375310) \\ 
15 & (1.110,-0.346) & (1.1497,-0.3353) & (1.14598003,-0.33019962) & 
(1.14702383,-0.332562329) \\ 
20 & (1.160,-0.307) & (1.1415,-0.3287) & (1.14512823,-0.33023791) & 
(1.14511731,-0.330140243) \\ 
25 & (1.141,-0.354) & (1.1478,-0.3298) & (1.14523860,-0.33015825) & 
(1.14523597,-0.330179581) \\ 
30 & (1.139,-0.312) & (1.1437,-0.3311) & (1.14522511,-0.33018993) & 
(1.14522395,-0.330182552) \\ 
35 & (1.157,-0.341) & (1.1458,-0.3290) & (1.14522377,-0.33017859) & 
(1.14522539,-0.330181124) \\ 
40 & (1.131,-0.325) & (1.1451,-0.3312) & (1.14522642,-0.33018226) & 
(1.14522527,-0.330181563) \\ 
45 & (1.158,-0.327) & (1.1448,-0.3294) & (1.14522458,-0.33018138) & 
(1.14522524,-0.330181440) \\ 
50 & (1.135,-0.337) & (1.1457,-0.3305) & (1.14522559,-0.33018133) & 
(1.14522527,-0.330181470) \\ 
55 & (1.149,-0.320) & (1.1446,-0.3300) & (1.14522512,-0.33018161) & 
(1.14522525,-0.330181465) \\ 
60 & (1.145,-0.340) & (1.1456,-0.3300) & (1.14522528,-0.33018135) & 
(1.14522526,-0.330181464) \\ 
65 & (1.140,-0.322) & (1.1449,-0.3304) & (1.14522527,-0.33018153) & 
(1.14522525,-0.330181466) \\ 
70 & (1.152,-0.334) & (1.1453,-0.3298) & (1.14522523,-0.33018143) & 
(1.14522526,-0.330181465) \\ 
75 & (1.137,-0.329) & (1.1452,-0.3304) & (1.14522528,-0.33018147) & 
(1.14522526,-0.330181466) \\ 
80 & (1.152,-0.327) & (1.1450,-0.3296) & (1.14522524,-0.33018146) & 
(1.14522526,-0.330181465) \\ 
85 & (1.140,-0.335) & (1.1454,-0.3302) & (1.14522527,-0.33018145) & 
(1.14522526,-0.330181465) \\ 
90 & (1.147,-0.323) & (1.1450,-0.3301) & (1.14522525,-0.33018147) & 
(1.14522526,-0.330181465) \\ 
95 & (1.146,-0.336) & (1.1453,-0.3300) & (1.14522526,-0.33018145) & 
(1.14522526,-0.330181465) \\ \hline
& \multicolumn{4}{c|}{$G_{A} =(1.14522526,-0.330181465) $}
\end{tabular}
}
\caption{Convergence of the continued fraction for first element of the
Green's matrix at $\Re \epsilon > 0$ with method B. The first, second, third
and fourth column contain approximants of the continued fraction with $%
w_n=w_+$ and without, with one-fold, with five-fold and with eight-fold
Bauer-Muir transform, respectively. For comparison we also give the exact
result. All the $G_{00}$ values are scaled with $10^2$.}
\end{table}

\end{document}